# Extension of the Wu-Jing equation of state (EOS) for highly porous materials: thermoelectron based theoretical model

Geng Huayun,[*] Wu Qiang, Tan Hua, Cai Lingcang and Jing Fuqian

Laboratory for Shock Wave and Detonation Physics Research, Southwest Institute of Fluid Physics, P. O. Box 919-102, Mianyang Sichuan 621900, People's Republic of China

A thermodynamic equation of state (EOS) for thermoelectrons is derived which is appropriate for investigating the thermodynamic variations along isobaric paths. By using this EOS and the Wu-Jing (W-J) model, an extended Hugoniot EOS model is developed which can predict the compression behavior of highly porous materials. Theoretical relationships for the shock temperature, bulk sound velocity, and the isentrope are developed. This method has the advantage of being able to model the behavior of porous metals over the full range of applicability of pressure and porosity, whereas methods proposed in the past have been limited in their applicability.

PACS: 64.30.+t, 62.50.+p, 61.43.Gt, 47.40.Nm

---

[*] E-mail: genhy@sohu.com.

## I. INTRODUCTION

The equation of state (EOS) of porous materials have been studied extensively in terms of theoretical models and experiments in the low-porosity region, where the shock temperature is several thousands of degrees Kelvin.[1-8] In this regime, the effect of thermoelectrons can be ignored. Among these theoretical models, the one proposed by Wu and Jing (W-J)[7] is more appropriate because it combined the $p-\alpha$ model[2] for the low pressure region with the unique W-J relation and allows calculation of the thermodynamic variables along isobaric paths; it is able to predict the whole Hugoniot path for a material. The W-J model has been compared to other analytical methods by Boshoff-Moster and Viljoen.[9]

However, the W-J EOS does not do a good job of predicting the behavior of highly-porous materials when the effect of thermoelectrons become important in the compression process.[10-15] The EOSs in this regime are much more complicated than those discussed above. The pressure as a function of specific volume becomes multivalued under these conditions as depicted by Hugoniot 2 in Fig.1.

The earliest attempt to develop an EOS model for highly porous materials was made by Trunin et al.[11,12] They extrapolated the thermodynamic variables from the solid Hugoniot to the porous Hugoniot along the tangent to the latter. In a certain sense, this model works well in the regime of porosity $m \leq 5$ and pressure $P \leq 100 GPa$, where $m=V_{00}/V_0$; $V_{00}$ and $V_0$ are the initial specific volumes of the porous and solid material, respectively. However, as an empirical method, this model appears too complex and is without a unified theoretical basis.

In 1997, Trunin proposed an advanced model in which the thermal capacity due to anharmonic vibrations of the crystal lattices was varied and the effect of thermoelectrons was considered.[13] This model predicts the experimental data better but is more complicated and artificial than his previous model. It has been criticized for using inexact experimental data of an expanded solid in the calculations.

A different EOS model for highly-porous materials was suggested by Gryaznov et al.[16,17] by assuming the shocked material is a non-ideal plasma; it is called the "non-ideal plasma model." It is applicable to the high temperature regime or thermoelectron regime discussed in this paper. The shocked porous material is treated as a mixture of electrons, atoms, and ions of different charges interacting with each other. The free energy of such a system is split into two parts: 1) the ideal-gas contribution of atoms, ions, and electrons and 2) the part responsible for inter-particle interactions. Obviously, this model suffers from being invalid in low-pressure, low-temperature regimes.

The developments in this paper are based on the idea that it is more natural to extrapolate the porous Hugoniot from the solid Hugoniot (or a gas Hugoniot) along isobaric paths rather than isochoric paths. The basis for this choice is easily understood with the aid of the illustration of Fig.1. A Mie-Grüneisen EOS, which is based on variations along an isochoric path, can be used to calculate porous Hugoniot 1 in the figure (e.g., point *d* on porous Hugoniot 1 can be extrapolated from point *a* on the solid Hugoniot). However, when the initial porosity is greater so that porous Hugoniot 2 is appropriate, this extrapolation can not be made. If one desired to extrapolate point *e* on this Hugoniot from the solid Hugoniot, the initial state would have to be determined first. Since there is no initial state on the solid Hugoniot corresponding to a specific volume of $V_2$, the extrapolation becomes difficult unless one determines an initial value artificially, as was done by Trunin. Even then it is difficult to determine whether the extrapolated point should be *e* or *f*. Trunin's methods[11, 13] have these disadvantages because they are based on the Mie-Grüneisen EOS. Gryaznov's method is also based on the similar idea as Mie-Grüneisen EOS and suffers from the same disadvantages. However, if one uses the W-J EOS that calculates along an isobaric path, these difficulties are eliminated and one can extrapolate the porous Hugoniots 1 and 2 directly (from point *c* to *d* and *e*, or from point *a* to *b*).

The purpose of this article is to extend the W-J EOS to the highly-porous regime, i.e., to develop an advanced model to predict the shock properties of porous materials to the regime of *T*<60000*K* and *m*<20 by calculating the thermodynamic variations along isobaric paths. In section Ⅱ, a new thermodynamic relationship for a near-free system is derived. Then, by making use of this relation, an EOS for free thermoelectrons in metals is developed. The complete formation of the extended model is presented in section III. Relationships, based on this model, are developed for calculating shock temperature, bulk sound velocity, and isentrope in sections IV and V. Comparisons of thermodynamic states calculated using this model with other models and experimental data are contained in a companion paper.

Through this whole article, we assume the W-J EOS relationship[7]

$$V_a - V_b = \frac{R}{P}(H_a - H_b), \tag{1}$$

or alternatively,

$$(\partial V/\partial H)_P = R/P, \text{ if } b \to a, \tag{1a}$$

is valid for any two states '*a*' and '*b*' on a *P-V* plot in general. Here the symbols *V*, *H*, *P* and *R* denote the specific volume, the specific enthalpy, the pressure, and the W-J EOS parameter of the material, respectively.

## II. THERMODYNAMIC EOS FOR THERMOELECTRONS

### A. Derivation of the new thermodynamic relation

Considering a system whose Hamiltonian is absolutely separable, namely,

$$H(x_i^\mu) = \sum_{i=1}^n H_i(x_i), \tag{2}$$

and
$$x_i \cap x_j = \phi, \quad (i \neq j), \tag{3}$$

where *H* is the system's Hamiltonian (in this article, unless otherwise specified, *H* denotes the enthalpy in general), $\phi$ is the null set, $x_i^\mu$ is the set of all canonical coordinates of the system as *i*=1,2,3…and $\mu$ =1,2…, and $x_i$ is the subset of $x_i^\mu$ just as $\mu$ =1,2,3…, the pressure equation of this system is[18]

$$PV = \kappa T \ln \Xi \tag{4}$$

in a grand ensemble. Here, $\Xi$ is the grand partition function that satisfies

$$\Xi = \sum_{N_i=0}^\infty \prod_{i=1}^n e^{-N_i \alpha_i} Z_{N_i} \tag{5}$$

with
$$Z_{N_i} = \frac{1}{N_i!(2\pi\hbar)^{3N_i}} \int e^{-\beta H_i} (dx_i)^{N_i}, \tag{6}$$

when the system is chemically-pure. Here $N_i$ is the particle number of subsystem *i*. Taking the conservation of particle number of each subsystem into account, namely, taking all $N_i$ fixed, we can rewrite Eq.(5) as

$$\Xi = \prod_i \Xi_i = \prod_{i=1}^n e^{-N_i \alpha_i} Z_{N_i}. \tag{7}$$

Therefore, the pressure equation of the whole system Eq.(4) becomes

$$PV = \kappa T \ln \Xi = \sum_i \kappa T \ln \Xi_i. \tag{8}$$

It is apparent by Eq.(8) that one can write the specific volume of the whole system as a summation of all specific-volume-increments due to the involved subsystems, namely,

$$V = \frac{1}{P} \sum_i \kappa T \ln \Xi_i = \sum_i V_i, \tag{9}$$

where
$$V_i = \frac{\kappa T}{P} \ln \Xi_i. \tag{10}$$

On the other hand, thanks to the decoupling of the subsystems, the presence of any other subsystems should have no influence on subsystem *i*. So, its pressure equation should still be

$$P_i V = \kappa T \ln \Xi_i. \tag{11}$$

Combining Eqs.(10) and (11) yields a new thermodynamic relation for decoupled systems:

$$P_i V = P V_i. \tag{12}$$

Eq.(12) works only when the pressure is not equal to zero since it is the denominator in Eq.(10).

**B. New equation of state for thermoelectrons**

Now, we would like to apply Eq.(12) to the case of free thermoelectrons to derive a new equation of state for metals. A metal crystal is a typical multiparticle system consisting of atomic nuclei and electrons. Though the equations of motion for this system are simple, it is impossible to solve the Schrödinger equations analytically. However, the equations can be divided into two parts: one for the electrons and the other for the nucleus, if use has been made of the Born-Oppenheimer approximation.[18, 19] For the electrons, it is convenient to assume they make up a free Fermi gas which moves in a square potential well with an infinite depth surrounded by the surfaces. Then, the Hamiltonian of this part is independent of that of the nucleus part, and vice versa. In this way, if the numbers of nucleus and electrons are conserved at the final states of compression, the requirements of Eq.(12) are all met and we have

$$P_e V = P V_e. \tag{13}$$

Here subscript $e$ denotes variables contributed by electrons.

Substituting Eq.(13) into the electron Mie-Grüneisen EOS $P_e = \gamma_e E_e / V$, where $E$ is the specific internal energy and $\gamma_e$ the electronic Grüneisen parameter, we get the specific volume of electrons

$$V_e = \gamma_e E_e / P. \tag{14}$$

In conventional treatments the cold energy of the electrons is always merged into the cold part of crystals. Therefore, what remains, $E_e$, is the internal energy of thermoelectrons and it just depends on temperature,

$$E_e = \beta T^2 / 2. \tag{15}$$

Correspondingly, the specific volume of thermoelectrons becomes

$$V_e = \gamma_e \beta T^2 / 2P. \tag{16}$$

Here, $\beta$ is the coefficient of electronic specific heat and can be evaluated by $\beta = \beta_0 (V/V_{0K})^{\gamma_e}$ in general, where $V_{0K}$ is the initial specific volume of the solid at zero-Kelvin. Obviously, the specific enthalpy of thermoelectrons can be obtained from Eqs.(15) and (16) as

$$H_e = E_e + PV_e = \frac{\gamma_e + 1}{2} \beta T^2. \tag{17}$$

## III. HUGONIOT EOS FOR HIGH-POROSITY MATERIALS

Wu and Jing made a helpful discussion about the equation of state for low-porosity materials in 1996.[7] They established an EOS model to extrapolate the porous Hugoniot from the solid one along isobaric paths; this works well when the porosity is low. However, in their model they included the enthalpy of thermoelectrons with other parts. This resulted in an awkward situation where the thermoelectronic enthalpy was controlled by the W-J EOS parameter for the crystal part, rather than having a separate part for the thermoelectrons. In cases involving highly porous materials, this difference is significant.

To extend their EOS model to a much wider range of applicability, we separate out the thermoelectronic enthalpy and put it under the control of its own W-J parameter. Consequently, the relevant EOSs for the solid and the corresponding porous materials can be written in the following forms, respectively:

$$V_h - V_{ne} = \frac{R_e}{P}(H - H_{ne}) \quad \text{and}$$

$$V_{ne} - V_x = \frac{R_c}{P}(H_{ne} - H_x) \quad \text{for the solid material;} \tag{18}$$

$$V_h' - V_{ne}' = \frac{R_e}{P}(H' - H_{ne}') \quad \text{and}$$

$$V_{ne}' - V_x' = \frac{R_c}{P}(H_{ne}' - H_x') \quad \text{for the porous material.} \tag{19}$$

Here the prime denotes the physical quantity for the porous material and subscripts $h$, $ne$, $x$ refer to the Hugoniot state, the Hugoniot state that has excluded the thermoelectrons' contribution, and the zero-Kelvin state at the same pressure, respectively. $R_c$ is regarded as an effective parameter for crystals with the same value both for the solid and porous materials under isobaric conditions. $R_e$ is the W-J parameter for thermoelectrons.

By the definition of specific enthalpy and Rankine-Hugoniot relations, we have for the porous material[10]

$$H_x' = PV_x' + E_x', \tag{20}$$

$$H_{ne}' = E_{00} + \frac{1}{2}P_1(V_{00} - V_1) + \frac{1}{2}P(V_1 + V_{ne}'), \tag{21}$$

$$H' = H_{ne}' + H_e' = H_{ne}' + \frac{\gamma_e + 1}{2}\beta T'^2, \tag{22}$$

and for the solid material

- 7 -

$$H_x = PV_x + E_x, \tag{23}$$

$$H_{ne} = E_0 + \frac{1}{2}P(V_0 + V_{ne}), \tag{24}$$

$$H = H_{ne} + H_e = H_{ne} + \frac{\gamma_e + 1}{2}\beta T^2 = E_0 + \frac{1}{2}P(V_0 + V_h), \tag{25}$$

where $E_{00}$ and $E_0$ are the initial specific internal energies of the porous and solid materials, respectively. The subscript 1 refers to the Hugoniot elastic limit (HEL) state of the porous material. Moreover, an additional assumption used in the following treatment is that the specific internal energy is the same for the porous material and the solid material under identical conditions of pressure and temperature, i.e., $E_{00} = E_0$ and $E_x' = E_x$. Combining Eqs.(18)-(25) results in a relationship between the porous Hugoniot and the solid Hugoniot under isobaric conditions,

$$V_h' = \frac{1-(R_c/2)}{1-(R_c/2)[1-(P_1/P)]}\left(V_h - \frac{3}{2P}\beta T^2\right) + \frac{R_c/2}{1-(R_c/2)[1-(P_1/P)]}\left[(V_1 - V_0) + \frac{P_1}{P}V_{00} + \frac{1-R_c}{(R_c/2)}(V_x' - V_x)\right] + \frac{(\gamma_e+1)R_e}{2P}\beta T'^2. \tag{26}$$

Eq.(26) is the extended Hugoniot EOS for highly porous materials we set out to develop in this paper. This Hugoniot EOS appears suitable for predicting the behavior full-range of the shocked highly-porous materials.

In our region of interest, e.g., pressures of $10GPa \leq P \leq 300GPa$, porosities of $m \leq 20$, and temperatures of $T \leq 60000K$, the following are true, $P >> P_1$ and $\beta T^2/P << V_h$. Using this, it is a good approximation to set $P_1 = 0$, $V_1 = V_{00}$, $V_x' = V_x$, $\gamma_e = 1/2$, and $\beta T^2/P \approx 0$. In addition, Eqs.(1), (16) and (17) give the W-J parameter for the thermoelectrons as $R_e = \frac{\gamma_e}{\gamma_e + 1}$. Thus, the Hugoniot relationship Eq.(26) can be rewritten as

$$V_h' = V_h + \frac{R_c}{2 - R_c}(V_{00} - V_0) + \frac{\beta T'^2}{4P}. \tag{27}$$

This form is much shorter than Eq.(26) and enables us to evaluate the shock temperature and sound

- 8 -

velocity more easily.

## IV.  DETERMINATION OF SHOCK TEMPERATURE

In order to evaluate the porous Hugoniot using Eq.(27), one must determine the shock temperature of the porous material. There are several methods to estimate the shock temperature for solid materials, but most of them do poorly in high-porosity cases since they do not consider the anomalous Hugoniot behavior at high porosities. For example, the method proposed by Walsh *et al.*[20] in 1957, though it works well in solid and near-solid cases, the differential equation of temperature is unsolvable in high-porosity cases because some parts of the highly porous Hugoniot are multivalued.

Considering that the Hugoniot relationship Eq.(27) contains the shock temperature of the porous material, the self-consistency of the model requires that this variable be evaluated within its frame. In this section, we will develop the Walsh method to meet this requirement by reestablishing it along an isobaric path.

Making use of some thermodynamic relations, one obtains the differential of specific enthalpy as

$$dH - VdP = \left(\frac{\partial H}{\partial T}\right)_P dT + \left(\frac{\partial H}{\partial P}\right)_T dP - VdP$$

$$= C_P dT - T\left(\frac{\partial V}{\partial H}\right)_P \left(\frac{\partial H}{\partial T}\right)_P dP = C_P dT - C_P T \frac{R}{P} dP. \qquad (28)$$

Here, use has been made of $(\partial H/\partial P)_T - V = -T(\partial V/\partial T)_P$ and $(\partial V/\partial H)_P = R/P$. On the other hand, the Rankine-Hugoniot relationship, along with $H = E + PV$, gives $dH - VdP$ as

$$dH - VdP = \frac{1}{2}\left(V_0 - V + P\frac{dV}{dP}\right)dP. \qquad (29)$$

By combining Eqs.(28) and (29), we acquire an ordinary differential equation of order one for shock temperature

$$\frac{dT}{dP} - \frac{R}{P}T = \frac{1}{2C_P}\left(V_0 - V + P\frac{dV}{dP}\right). \qquad (30)$$

This equation of shock temperature is universal in nature, and is applicable to both the solid and porous materials. Although it is similar to the primary Walsh equation

$$\frac{dT}{dV} + \frac{\gamma}{V}T = \frac{1}{2C_V}\left[P + (V_0 - V)\frac{dP}{dV}\right], \tag{31}$$

these is an essential difference. The former calculates along isobaric paths and the latter along isochoric paths. Consequently, Eq.(30) can be used to calculate the shock temperature of porous materials in association with the Hugoniot relations Eq.(26) or (27) but Eq.(31) cannot. The underlying multivalued nature of the high-porosity Hugoniot $P(V)$ makes the solution of Eq.(31) unobtainable.

By rewriting Eq.(30) so it applies to porous materials cases, one has

$$\frac{dT'}{dP} - \frac{R}{P}T' = \frac{1}{2C_P}\left(V_{00} - V_h' + P\frac{dV_h'}{dP}\right). \tag{32}$$

Here $V_{00}$ is the initial specific volume of the porous material. The derivative of the specific volume of the porous material with respect to pressure can be derived from Eq.(27) as

$$\frac{dV_h'}{dP} = \left(1 + \frac{\beta T'^2}{8PV_h}\right)\frac{dV_h}{dP} + \left[2(V_{00} - V_0)/(2 - R_c)^2\right]\frac{dR_c}{dP} + \frac{\beta T'}{2P}\frac{dT'}{dP} - \frac{\beta T'^2}{4P^2}. \tag{33}$$

Here use has been made of $\beta = \beta_0\left(\frac{V_h}{V_{0K}}\right)^{1/2}$, and has taken into account the fact that the shock Hugoniot depends only on pressure and temperature, so the other variables in Eq.(27) are just temporary symbols and must be replaced by the corresponding expressions of pressure and temperature when conducting the derivation. The pressure expression of the specific volume of the shocked solid material can be obtained by the solid Hugoniot relation $P = \rho_0 C_0^2(1 - \rho_0 V_h)/(1 - \lambda(1 - \rho_0 V_h))^2$, where $\rho_0 (=1/V_0)$, $C_0$ and $\lambda$ are the initial density, the sound velocity at normal state, and a material parameter coming from the relationship of shock wave velocity and particle velocity of the shocked material, respectively. The expression of $R_c$ is similar to that of the Mie-Grüneisen parameter, i.e., $R_c = \left(\frac{d\ln\Theta_D}{d\ln V}\right)_T\left(\frac{d\ln V}{d\ln P}\right)_T$, here $\Theta_D$ is the Debye temperature. The details can be found in the companion paper (Ref.21).

On the other hand, the parameter $R$ in Eq.(32) satisfies $(\partial V/\partial H)_P = R/P$. As we know, $H = H_{ne} + H_e$, which results in

- 10 -

$$\frac{P}{R} = \left(\frac{\partial H}{\partial V}\right)_P = \left(\frac{\partial H_{ne}}{\partial V}\right)_P + \left(\frac{\partial H_e}{\partial V}\right)_P = \frac{P(R_c + R_e)}{R_c R_e}, \tag{34}$$

thus
$$R = \frac{R_c R_e}{R_c + R_e} = \frac{R_c}{3R_c + 1}, \tag{35}$$

for $R_e = 1/3$.

Similarly, the specific heat at constant pressure in Eq.(32), which is defined by $C_P = (\partial H/\partial T)_P$, becomes

$$C_P = (\partial H/\partial T)_P = (\partial H_{ne}/\partial T)_P + (\partial H_e/\partial T)_P = C_{Pc} + C_{Pe}. \tag{36}$$

By Eq.(17) we have $C_{Pe} = (\gamma_e + 1)\beta T = 3\beta T/2$. The crystal part $C_{Pc}$, due to the influence of anharmonic vibrations of the lattices, decreases as the shock temperature increased. However, the relative ratio of its value, compared with the specific heat at constant volume, satisfies $C_{Pc}/C_{Vc} \approx 1$ when temperature trends to zero-Kelvin and $C_{Pc}/C_{Vc} \to 5/3$ when temperature approaches infinity, so we can assume that it is appropriate to let $C_{Pc}/C_{Vc} \approx const.$, at least for the first order approximation. Consequently, the anharmonic specific heat at constant pressure becomes

$$C_{Pc} = C_{Vc}C_{P0}/C_{V0} = C_{P0}\left[1 + (1+Z)^{-2}\right]/2, \tag{37}$$

since $C_{Vc}/C_{V0} \approx \left[1 + (1+Z)^{-2}\right]/2$.[10, 22] Here subscript 0 refers to a normal state, parameter $Z = lR_g T/\mu C_x^2$ is called the solid irrelevance where $R_g$ is the universal constant for gas, and the other parameters are as follows: $\mu$ is the mole mass, $C_x$ is the mean velocity of elastic waves, and $l$ is the anharmonic parameter.

Consequently, the final form of Eq.(32) becomes

$$\frac{dT'}{dP} - \frac{R_c/(3R_c+1)}{P}T' = \frac{1}{2C_P}\left(V_{00} - V_h' + P\frac{dV_h'}{dP}\right), \tag{38}$$

$$C_P = C_{P0}\left[1 + (1+Z)^{-2}\right]/2 + 3\beta T'/2. \tag{39}$$

Numerically solving this ordinary differential equation of order one for shock temperature (which is quite easy if one makes use of commercially available computing tools such as MATLAB, *etc*.), one obtains the shock temperature directly.

## V. ISENTROPE AND SOUND VELOCITY

Besides the shock temperature, sound velocity at tens or hundreds of Giga-Pascals (GPa) of pressure is also an important parameter to investigate since it provides information relating to the constitutive relations or equation of state of materials. One can learn important things about shocked materials (such as shock-induced melting) by measuring the sound velocities along shock Hugoniots. Moreover, sound velocity is also an important parameter to estimate the dynamic response of materials such as spallation, fragmentation, etc. Thus, evaluating the sound velocity on the basis of the extended Hugoniot relation Eq.(27) for porous materials is necessary. However, the same situation (as in the temperature calculation) exists in this case if the bulk sound velocity is defined in the conventional way, although there is no need for integration to obtain it. The method given by Walsh, which was based on the Mie-Grüneisen EOS,[10, 20] is an example of this. According to Walsh's theory, the slope of an isoentrope through a certain point on a shock Hugoniot is

$$\frac{\partial P_s}{\partial V} = -\frac{\gamma P_h}{2V} + \frac{\partial P_h}{\partial V}[1 - \frac{\gamma(V_0 - V)}{2V}], \tag{40}$$

and the bulk sound velocity can be calculated from $C^2 = -V^2 \frac{\partial P_s}{\partial V}$. For porous materials, this formula should be rewritten as

$$\left(\frac{\partial V_s}{\partial P}\right)^{-1} = -\frac{\gamma P}{2V_h'} + \left[1 - \frac{\gamma(V_{00} - V_h')}{2V_h'}\right] / \frac{\partial V_h'}{\partial P}, \tag{41}$$

and

$$\frac{\partial V_h'}{\partial P} = \left(1 + \frac{\beta T'^2}{8PV_h}\right)\frac{dV_h}{dP} + \left[2(V_{00} - V_0)/(2 - R_c)^2\right]\frac{dR_c}{dP} - \frac{\beta T'^2}{4P^2}. \tag{42}$$

In the derivation of Eq.(42), use has been made of the fact that the Hugoniot relation Eq.(27) depends only on temperature and pressure, so $\frac{\partial V_h'}{\partial P} = \frac{dV_h'}{dP} - \frac{\partial V_h'}{\partial T'}\frac{dT'}{dP} = \frac{dV_h'}{dP} - \frac{\beta T'}{2P}\frac{dT'}{dP}$.

Though Eq.(41) provides good predictions in the case of solid and near-solid porous materials, there is a flaw in that it becomes invalid when

$$\left(\frac{\partial V_h^{'}}{\partial P}\right)^{-1}\left[2V_h^{'} - \gamma\left(V_{00} - V_h^{'}\right)\right] > \gamma P. \tag{43}$$

Unfortunately, most Hugoniots of high-porosity materials cannot meet this requirement.

In order to use the Walsh method, it would be necessary to rework Eq.(41) so that it applied to isobaric conditions. We prefer to begin with the W-J relation

$$V_s - V_h^{'} = \frac{R}{P}\left(H_s - H^{'}\right). \tag{44}$$

Here the specific volume $V_s$ of the isentrope is not marked with a prime because the isentrope is a thermodynamic function so it should be porosity free. Using the thermodynamic relation $\left(\partial H/\partial P\right)_s = V_s$, the porous Rankine-Hugoniot relations, and taking the local derivative of Eq.(44) with respect to pressure directly, we have

$$\frac{\partial V_s}{\partial P} = \left(1 - \frac{R}{2}\right)\frac{\partial V_h^{'}}{\partial P} - V_h^{'}\left(\frac{R-2}{2P} + \frac{dR}{RdP}\right) - \frac{R}{2P}V_{00} + V_s\left(\frac{R-1}{P} + \frac{dR}{RdP}\right). \tag{45}$$

Here $R$ is the total W-J parameter and determined by Eq.(35).

Setting $V_s = V_h^{'}$, Eq.(45) gives the sound velocity under shock conditions as $C^2 = -V_h^{'2}\left(\partial V_s/\partial P\right)_h^{-1}$ and

$$\left(\frac{\partial V_s}{\partial P}\right)_h = \left.\frac{\partial V_s}{\partial P}\right|_{V_s = V_h^{'}} = \left(1 - \frac{R}{2}\right)\frac{\partial V_h^{'}}{\partial P} + \frac{R}{2P}\left(V_h^{'} - V_{00}\right); \tag{46}$$

i.e., this is the high-pressure sound velocity for a porous material that we set out to obtain in this section. However, it is constrained by

$$\left(\frac{2-R}{R}\right)\frac{\partial V_h^{'}}{\partial P} \leq \frac{V_{00} - V_h^{'}}{P}, \tag{47}$$

which is a much looser condition than Eq.(43) so most porous materials will satisfy it.

## VI. SUMMARIES

A thermodynamic equation of state for thermoelectrons has been derived in this article based on the W-J EOS model. It is appropriate for analyzing the thermodynamic conditions along isobaric paths, rather than along isochoric paths which would be the case if the well-known Mie-Grüneisen EOS were used. By using this EOS, an EOS model for predicting the Hugoniot of porous materials (with the corresponding solid Hugoniot as reference) is developed, which has the advantage of full-pressure and full-porosity range applicability. Models proposed in the past do not cover the full range.

Methods for calculating the shock temperature and the bulk sound velocity were developed, based on the general W-J relation, so that a set of equations consistent with isobaric thermodynamic states can be obtained. These have replaced their counterparts, which were based on the Mie-Grüneisen relation and were derived along isochoric paths. This new model has been used to calculate material properties under shock conditions over a wide range of pressure, porosity, and temperature for a number of materials; the results of these calculations are given in the companion paper (Ref.21).

**Acknowledgements**

This study was financially supported by the National Natural Science Foundation of China under Grant No. 19804010 and Science and Technology Foundation of CAEP under Grant No. 980102.

Figure caption:

Fig.1. Comparison of the methods for calculating the thermodynamic states along isochoric paths (Mie-Grüneisen EOS) and those along isobaric paths (W-J EOS). For a normal porous Hugoniot (m<3, depicted by porous Hugoniot 1), the two methods are both valid since it is possible to extrapolate from the solid Hugoniot to the porous Hugoniot. For a highly-porous Hugoniot (3<m<20, depicted by porous Hugoniot 2), the Mie-Grüneisen EOS is invalid for making the extrapolation because it lacks an initial state to start from and also the Hugoniot is multivalued (e.g., points *e* and *f* are at the same volume $V_2$). However, the W-J EOS is valid because it does not suffer from these difficulties.

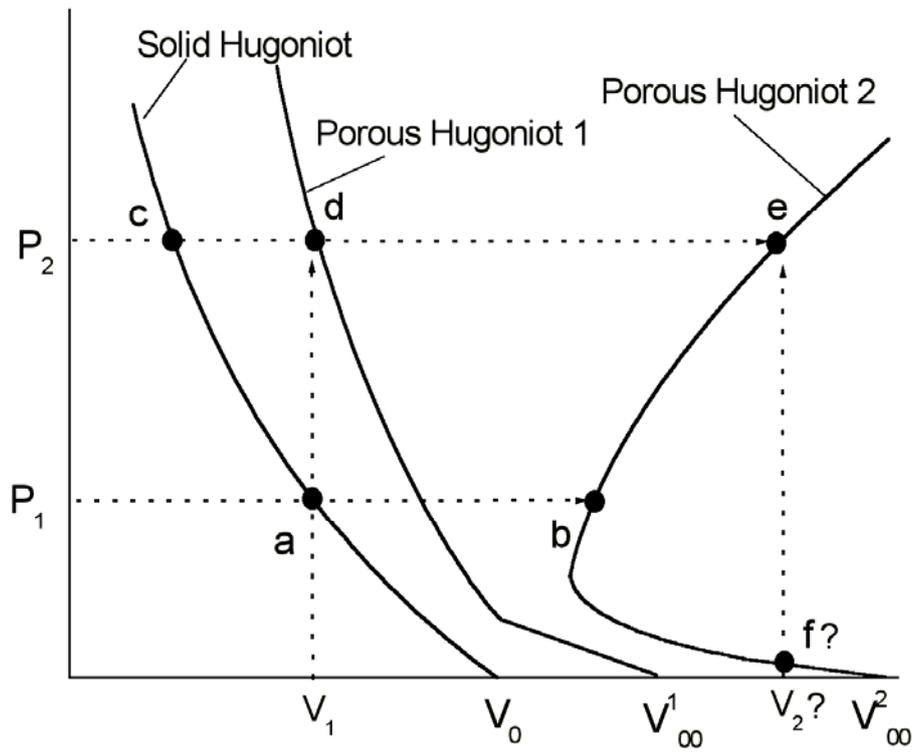

Figure 1